\documentclass[aps,tightenlines,showpacs]{revtex4}
\usepackage[dvips]{graphicx}
\usepackage[english]{babel}
\usepackage[T1]{fontenc}
\usepackage{mathrsfs}
\usepackage[tbtags]{amsmath}
\usepackage{amssymb}
\usepackage{amsxtra}
\usepackage{amsopn}
\usepackage{latexsym}
\usepackage[mathcal]{eucal}

\newcommand{\BE}{\begin{equation}}
\newcommand{\EE}{\end{equation}}
\newcommand{\BA}{\begin{eqnarray}}
\newcommand{\EA}{\end{eqnarray}}

\begin{document}

\title{Non-perturbative effective model
for the Higgs sector of the Standard Model}

\author{Fabio Siringo}
\author{Luca  Marotta}
\affiliation{Dipartimento di Fisica e Astronomia, 
Universit\`a di Catania,\\
INFN Sezione di Catania and CNISM Sezione di Catania,\\
Via S.Sofia 64, I-95123 Catania, Italy}

\date{\today}
\begin{abstract}
A non-perturbative effective model is derived 
for the Higgs sector of the 
standard model, described by a simple scalar theory.
The renormalized couplings are determined by the derivatives
of the Gaussian Effective Potential that are known to be 
the sum of infinite bubble graphs contributing to the vertex functions.
A good agreement has been found with strong coupling lattice simulations
when a comparison can be made. 

\end{abstract}


\maketitle

\section{introduction}

The nature and eventual existence of the Higgs Boson is one of
the major problems in particle physics. While direct searches  
already provide a lower bound above 100 GeV \cite{lep} for 
the Higgs mass, electroweak precision
measuraments still suggest a light Higgs, unless new physics 
should be found at the TeV scale by the LHC: in that case
a heavy Higgs would require the study of non-perturbative effects
on the effective theory which describes the Higgs sector.

On the other hand a light Higgs is surely consistent with a
perturbative treatment of the Standard Model (SM), but does not rule out
a large self-coupling of the Higgs sector, since a light Higgs
mass has also been 
predicted\cite{siringo_var, siringo_light, siringo_su2, ibanez}
in the strong coupling regime
by non-perturbative methods. 
Thus, even for a light Higgs,
some important non-perturbative effects may come into play.

Moreover, even if the self coupling were weak, a strong Yukawa coupling
is expected for the top quark, and non-perturbative effects cannot be
ruled out in the resulting Higgs-top coupled theory. 
It has been recently shown\cite{siringo_su2} 
that the weak couplings of the full $SU(2)\times U(1)$ 
gauge theory play a  very negligible role in the Higgs sector, and that the simple
self-interacting $\lambda \phi^4$ scalar theory\cite{siringo_var, siringo_light} 
yields the same predictions of the full non-Abelian gauge theory.
Thus we believe that the simple scalar theory, and its
scalar-fermion extension\cite{kuti2}
still represent a valid starting point in the study of
the non-perturbative features of the Higgs sector. 

The scalar theory has been extensively 
studied in the past, but few truly non-perturbative treatments
have been reported, and most of them have failed in the
attempt to give a fully consistent interacting theory
independent of any regularization scheme. More recently
the triviality of the theory has been generally accepted 
and nowadays the SM is regarded as an effective low energy
model: it does not need to be weakly coupled nor renormalizable,
but it is supposed to be only valid up to some energy scale.
An energy cut-off may be used as a regulator of the diverging
integrals, and must be left finite in order to avoid to front
an useless non-interacting trivial theory.
In this framework there has been a renewed interest in the 
non-perturbative behaviour of scalar theories with a large but finite
energy cut-off\cite{kuti1,kuti2}. Lattice simulations are the most reliable
non-perturbative approaches: even if they fail to provide any analytical
description themselves, they can be regarded as the best benchmark 
for testing approximate analytical tools. Unfortunately there are not
many of such tools, and in the case of a single scalar field $1/N$ expansions are
ruled out.

The Gaussian Effective Potential
(GEP)\cite{schiff,rosen,barnes,kuti,chang,weinstein,huang,bardeen,peskin,stevenson}
is a simple variational tool which has been often used for describing the 
spontaneous symmetry breaking in the framework of the scalar theory for
particle physics and condensed matter\cite{camarda,marotta1,marotta2}.
Moreover, as it has been already pointed out\cite{ibanez1,ibanez2}, the GEP contains
information about the renormalized one-particle-irreducible (1PI) n-point
functions: the derivatives of the GEP are a variational estimate of
the 1PI functions at zero external momenta. Thus a direct expansion of the GEP
around the broken-symmetry vacuum can be regarded as a an effective low-energy
model with the derivatives that act as renormalized couplings.

In this paper we study the emerging effective model and discuss the meaning
of the couplings in terms of infinite sums of special classes of graphs. 
In fact the 
couplings are known\cite{ibanez1,ibanez2,kovner1,kovner2,consoli85,consoli93} 
to be given by the infinite sum of all the
bubble graphs which can be drawn as perturbative corrections for 
the variationally optimized Gaussian Lagrangian. At variance with previous work
the derivation of the effective model and its renormalization is carried out at
a finite cut-off exactly, without any further approximation that would
require or assume a very large cut-off which should eventually be sent to infinite.
Thus the calculation is in the spirit of recent lattice simulations\cite{kuti1,kuti2}, 
and can be compared to such numerical calculations 
while providing an analytical non-perturbative low-energy effective model.
When a direct comparison can be made, we find a very good agreement with lattice
data\cite{oldMC1} gaining confidence on the reliability of the variational method. 

It is worth pointing out that, while the method is not new by itself, no previous
attempt had been made to compare the predictions of the finite cut-off effective
model with the available lattice data. For instance we derive a very simple
analytical expression for the transition point that fits the lattice data very
well up to very large couplings. We think that the method could 
be extended and used for an analytical description of the Higgs-top model 
which has been recently addressed by lattice simulation\cite{kuti2}

Another important issue is the existence of an intermediate mass which plays
the role of a variational parameter and appears as an intermediate energy scale:
quite smaller than the cut-off and still rather larger than the physical
masses. Since all internal lines in bubble graphs are evaluated with such intermadiate
mass, then the resulting n-point functions have a very weak dependence on the
external momenta at low energy, and they can be regarded as constant renormalized
couplings. In other words the physical low-energy amplitudes can be derived 
at tree-level by an effective lagrangian whose couplings already contain
the sum of infinite bubble graphs. Moreover, the non-perturbative nature
of the method does not require the self-coupling to be small.

The paper is organized as follows: 
in section II the effective model is defined for a self-interacting scalar field;
in section III the couplings are recovered as the sum of bubble
expansions around the Gaussian ansatz; in section IV the problem of field
renormalization is addressed and the nature of the critical point
is studied; in section V the phenomenological content of the effective model
is discussed; some final remarks and directions for future work
are reported in section VI.

\section{The variational effective model}

Neglecting couplings to other fields\cite{siringo_su2}, 
the Higgs sector of the Standard Model 
can be described 
by a simple scalar theory: in the Euclidean formalism
the Lagrangian reads
\BE
{\cal L}=\frac{1}{2}\partial^\mu\phi\>\partial_\mu\phi+\frac{1}{2}
m_B^2\phi^2+\frac{1}{4!}\lambda_B\phi^4.
\EE
Denoting by $\varphi$ the expectation value of the scalar field
$\varphi=<\phi>$, and by $h$ the Higgs field $h=\phi-<\phi>$, the
Lagrangian can be split as
\BE
{\cal L}={\cal L}_{GEP}+{\cal L}_{int}
\label{Ldef}
\EE
where
\BE
{\cal L}_{GEP}=\frac{1}{2}\partial^\mu h\>\partial_\mu h
+\frac{1}{2} \Omega^2 h^2
\label{Lgep}
\EE
and ${\cal L}_{int}={\cal L}-{\cal L}_{GEP}$.
The GEP can be recovered\cite{stevenson}
as the first order effective potential
for the free theory described by ${\cal L}_{GEP}$ in presence
of the interaction ${\cal L}_{int}$. The mass $\Omega$ of the free
theory is then determined by requiring that for any value
of the average $\varphi$ the effective potential
is at a minimum. 
A trivial calculation of the first order effective potential yields
\BE
V_{GEP}(\varphi)=\frac{1}{2} m_B^2\varphi^2
+\frac{1}{2} m_B^2 I_0(\Omega)
+\frac{\lambda_B}{4!}\varphi^4
+\frac{\lambda_B}{4}\varphi^2I_0(\Omega)
+\frac{\lambda_B}{8} \left[I_0(\Omega)\right]^2
-\frac{1}{2}\Omega^2 I_0(\Omega)+I_1(\Omega)
\label{GEP}
\EE
where 
the Euclidean integrals $I_0$, $I_1$ are defined as
\BE
I_0(X)=\int_\Lambda \frac{d_E^4k}{(2\pi)^4}\frac{1}{k^2+X^2}
\label{I0}
\EE
\BE
I_1(X)=\frac{1}{2}\int_\Lambda \frac{d_E^4k}{(2\pi)^4}\log(k^2+X^2).
\label{I1}
\EE
Here the symbol $\int_\Lambda$ means that the integrals are regularized
by insertion of a cut-off $\Lambda$ so that $k<\Lambda$:
according to the well known triviality of the $\lambda\phi^4$ 
theory the Higgs sector is regarded as an effective model with a high energy scale
$\Lambda$ which plays the role of a further free parameter\cite{kuti2}.
The GEP is given by the effective potential $V_{GEP}$ in Eq.(\ref{GEP}) 
provided that the mass parameter $\Omega$ is regarded as an implict function
of $\varphi$, defined by the minimum condition (gap equation)
$\partial V_{GEP}/\partial \Omega=0$ which reads\cite{stevenson}
\BE
\Omega^2=m_B^2+\frac{\lambda_B}{2}\varphi^2+\frac{\lambda_B}{2} I_0(\Omega).
\label{gap}
\EE
The phenomenological broken-symmetry minimum of the GEP occurs at 
$\varphi=\varphi_0$ where the partial derivative of the GEP vanishes.
By insertion of the gap equation Eq.(\ref{gap}) the derivative of $V_{GEP}$ reads
\BE
\frac{\partial V_{GEP}}{\partial \varphi^2}=\frac{1}{2}
\left(\Omega^2-\frac{1}{3}\lambda_B\varphi^2\right)
\label{V2}
\EE
and it vanishes at 
\BE
\varphi_0^2=\left[\frac{3\Omega^2}{\lambda_B}\right]_{\varphi=\varphi_0}.
\label{tree}
\EE
The known phenomenology of the Standard Model requires that $\varphi_0=v=247$ GeV, and
then Eq.(\ref{tree}) gives the mass parameter $\Omega_0=\Omega\vert_{\varphi=\varphi_0}$ 
at the minimum 
as a function of the
bare self-coupling $\lambda_B$. At the minimum point $\varphi=\varphi_0$
the gap equation Eq.(\ref{gap}) can be
satisfied by a proper choice of the free parameter $m_B^2$, and the theory has one
only free parameter, i.e. the bare self-coupling $\lambda_B$ 
(besides the cut-off $\Lambda$). We must stress that $\Omega$ is a simple variational 
parameter and there is no reason to believe that its value has any physical relevance.
In fact the mass $\Omega$ is the mass of a free particle described
by the unperturbed Lagrangian ${\cal L}_{GEP}$, while the true Higgs mass comes out
from the 1PI 2-point function $\Gamma_2$ which is the infinite sum of 2-point graphs that
arise from the interaction ${\cal L}_{int}$. Of course a comparison of
Eq.(\ref{tree}) with the tree level
perturbative Higgs mass $M_h^2=\lambda_B v^2/3$ 
tells us that at the broken symmetry vacuum
$\varphi=\varphi_0=v$, and for a small bare coupling $\lambda_B\ll 1$,
the true mass is $M_h\approx \Omega_0$ as we expected since 
the residual interaction in ${\cal L}_{int}$ becomes very small. 

In general, the exact quantum effective action $\Gamma[\varphi]$ can be written as an
expansion in powers of $(\varphi-\varphi_0)$ around the vacuum expectation value
$\varphi_0$
\BE
\Gamma[\varphi]=\sum_n\frac{1}{n!}\prod_{i=1}^n 
\left[\int\frac{d^4p_i}{(2\pi)^4} \left(\varphi(p_i)-\varphi_0(p_i)\right)\right]
\Gamma_n(p_1,\dots,p_n)
\label{effaction}
\EE
where the functional derivatives $\Gamma_n$ are the exact 1PI n-point functions.
For the theory described by the Lagrangian ${\cal L}={\cal L}_{GEP}+{\cal L}_{int}$,
the n-point functions are the sum of infinite n-point graphs where the vertices are
read from the interaction ${\cal L}_{int}$ and the free propagator 
$G_0(p)=(p^2+\Omega_0^2)^{-1}$ arises from the optimized free-particle Lagrangian
${\cal L}_{GEP}$ with $\Omega$ set at the vacuum value $\Omega_0$
according to Eq.(\ref{tree}). 
Any physical amplitude may be evaluated as a sum of connected
tree graphs with vertices provided by the n-point functions $\Gamma_n$ which play
the role of renormalized couplings\cite{weinberg2}. 

Now suppose, as it turns out to be the case, that the variational parameter
$\Omega_0$ is quite large compared to the physical Higgs mass $M_h$. Then 
the n-point functions have a very weak dependence on the external momenta as
far as these are small or at least $p_i\approx M_h$. An approximate 
low energy effective model can be recovered by taking these couplings at their
zero momentum value 
$\Gamma_n\approx\Gamma_n(0)=\Gamma_n (0,\dots,0)$ and going back to the direct space
where an effective Lagrangian can be written as
\BE
{\cal L}_{eff}=\frac{1}{2}\partial^\mu h\>\partial_\mu h
+\frac{1}{2} \left[\Omega_0^2-\Gamma^\prime_2(0)\right] h^2
-\frac{1}{3!}\Gamma_3(0) h^3
-\frac{1}{4!}\Gamma_4(0) h^4+\cdots
\label{efflag}
\EE
having denoted by $\Gamma^\prime_2$ the sum of first and higher order contributions to
the 1PI 2-point function, i.e. all the terms except the zeroth-order one
which according to Eq.(\ref{Lgep}) is equal to $-\Omega_0^2$. 
The effective Lagrangian in Eq.(\ref{efflag})
describes a scalar Higgs field with a renormalized
mass $M_R^2=-\Gamma_2(0)=\Omega_0^2-\Gamma^\prime_2(0)$ and renormalized couplings
$g_R=-\Gamma_3(0)$, $\lambda_R=-\Gamma_4(0)$, and so on.
The advantage of the effective Lagrangian in Eq.(\ref{efflag}) is that any low energy
amplitude may be evaluated at tree level, neglecting all loops that have been
already summed up to all orders in the renormalized couplings. In other words the 
mass $M_R$ can be regarded as the true mass $M_h$ of the Higgs boson, and the couplings
$g_R$, $\lambda_R$ are related to the phenomenological scattering amplitudes of
the particle.

Unfortunately we do not have the exact quantum effective action $\Gamma[\varphi]$, but
we can extract a variational estimate of the couplings $\Gamma_n(0)$ from the GEP.
In fact for constant background fields $\varphi$, $\varphi_0$ the effective
action becomes the opposite of the effective potential, and if we set
$\varphi(p)=(2\pi)^4\delta^4(p)\varphi$ and 
$\varphi_0(p)=(2\pi)^4\delta^4(p)\varphi_0$ in Eq.(\ref{effaction}) then the
exact effective potential reads
\BE
V_{eff}(\varphi)=-\sum_n\frac{1}{n!} \Gamma_n(0)\left(\varphi-\varphi_0\right)^n
\label{vexp}
\EE
which is an expansion of the exact effective potential around $\varphi=\varphi_0$.
Thus the renormalized couplings can be calculated by the simple derivatives
of the effective potential as
\BE
\Gamma_n(0)=-\left[\frac{d\>^n V_{eff}}{d\varphi\>^n}\right]_{\varphi=\varphi_0}.
\label{derivatives}
\EE
On the other hand the GEP in Eq.(\ref{GEP}) is a variational approximation to
the exact effective potential $V_{eff}$ and we can evaluate a set of approximate 
couplings as
\BE
\Gamma_n(0)\approx-\left[\frac{d\>^n V_{GEP}}{d\varphi\>^n}\right]_{\varphi=\varphi_0}.
\label{gamma}
\EE
Insertion in Eq.(\ref{efflag}) provides a simple way to perform non-perturbative
low-energy calculations by the simple evaluation of connected tree graphs.
In spite of the approximations, in the strong-coupling 
regime of the Higgs sector
the predictions of this effective model are expected to be 
more reliable than perturbative calculations. 
In fact the approximate renormalized couplings in
Eq.(\ref{gamma}) are known to be the sum to all orders 
of bubble graphs for the vertex functions\cite{kovner1,kovner2,consoli93}.
Moreover the derivatives of the GEP in Eq.(\ref{gamma})
were shown to be a genuine variational approximation 
for the n-point functions\cite{kovner1,kovner2,ibanez1}
and were evaluated by several authors\cite{ibanez1,consoli85,consoli93}
in the past. Unfortunately such derivations strongly depend on the special
regularization scheme and on a series of approximations which only
make sense for a very large energy cut-off that should be eventually sent
to infinite. In more recent years, the general consensus on the triviality
of the scalar theory and the failure of any attempt to build a meaningful
model with an infinite cut-off, has changed our view of the standard model, 
and the more modest aim of an effective model has been generally accepted 
as a reasonable compromise\cite{kuti2}. In this framework it would be 
desirable to discuss the variational n-point functions as defined
in Eq.(\ref{gamma}) exactly, without any further approximation, at a
given cut-off $\Lambda$ which is supposed to be large but not too large,
and plays the role of a physical parameter that points to the energy scale
where new physics should become relevant. On the other hand, the resulting
effective model could be easily compared to lattice calculations where a
natural cut-off is supplied by the lattice spacing. In the next section
we derive the explicit expressions for the variational n-point functions
from the GEP, through Eq.(\ref{gamma}) and by direct sum of the 
equivalent bubble expansions.

\section{Bubble Expansions}

The best way to calculate the approximate couplings $\Gamma_n(0)$ 
is by derivatives of the GEP\cite{ibanez2}.
However it is instructive to see that the variational approximation Eq.(\ref{gamma})
is equivalent to the sum of all the tree bubble graphs that 
can be drawn for the vertex functions. By tree bubble graph we mean any
1PI graph which only contains chains of bubble insertions that do not make any
loop: in other words the external lines of a graph become 
disconnected whenever two lines belonging
to the same loop are cut. This set of graphs is a sub-class of the two particle
point reducible graphs\cite{verschelde}. For instance in Fig.1 the 7-loop graph (a)
is a tree bubble graph while the 8-loop graph (b) is not since its bubble chain 
makes a loop. Both of them are two particle point reducible.

\begin{figure}[ht]
\begin{center}
(a){\label{fig:subfig:a}\includegraphics[scale=0.5,bb=117 426 407 688,clip]{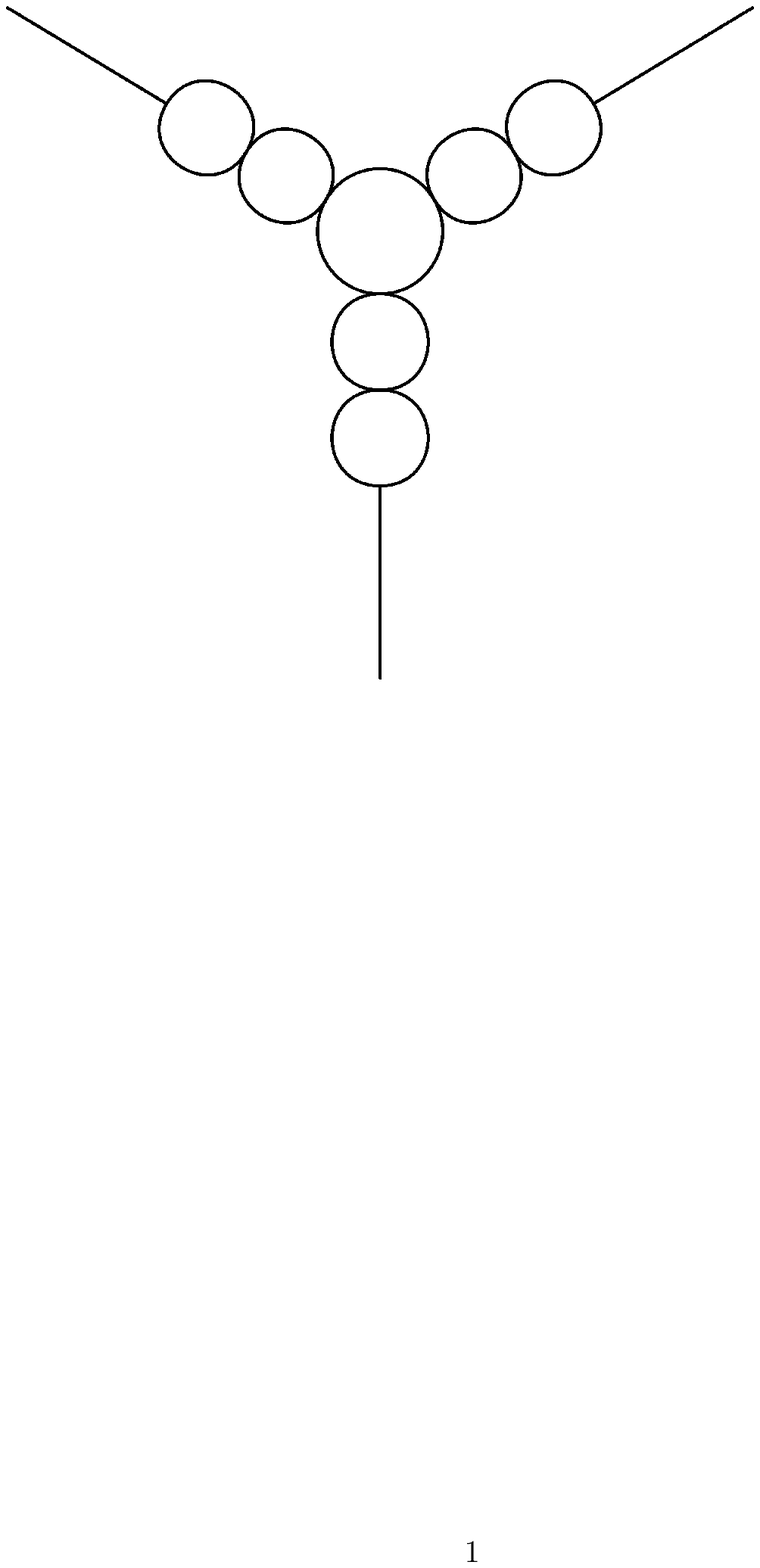}}
\hspace{1cm}
(b){\label{fig:subfig:b}\includegraphics[scale=0.5,bb=117 426 379 717,clip]{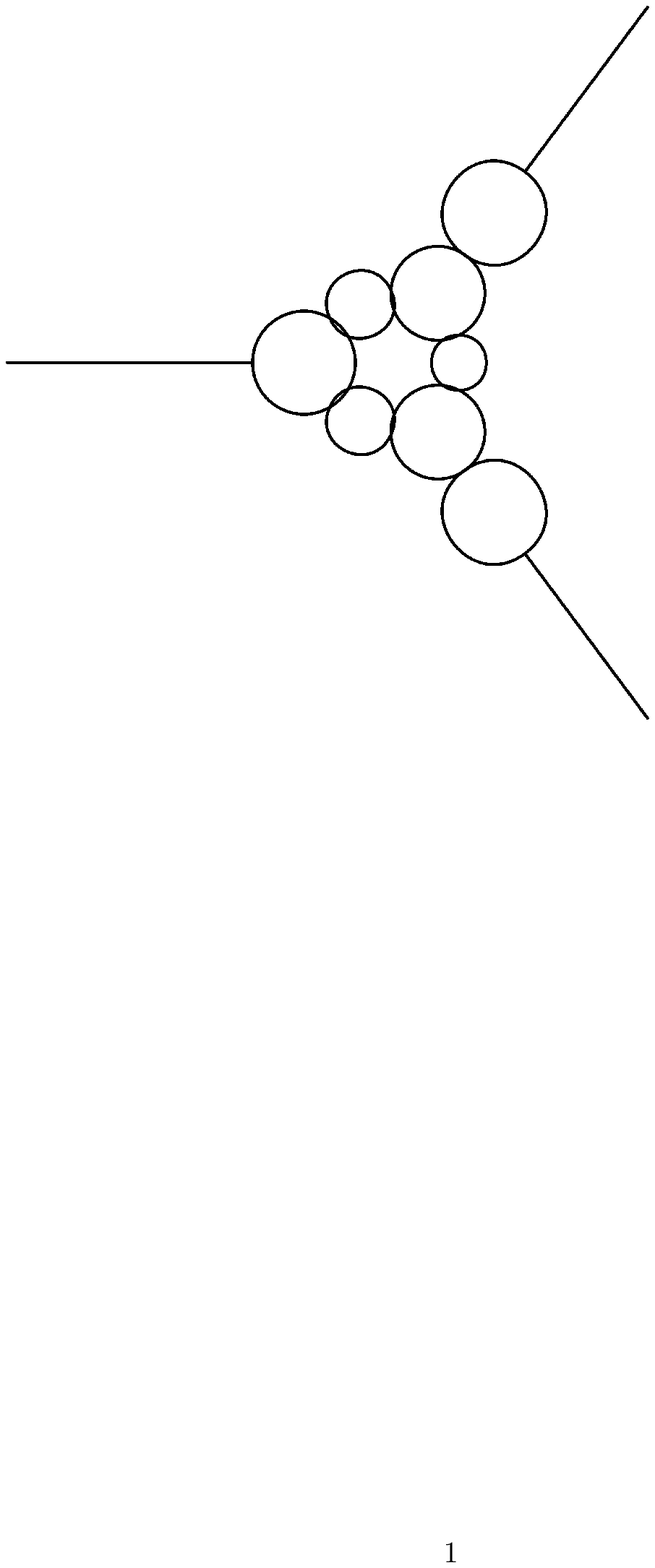}}
\renewcommand{\figurename}{Fig.}
\caption{\label{Fig1} Some two particle point reducible graphs. The 7-loop
graph (a) is a tree bubble graph as it becomes disconnected whenever two lines
belonging to the same loop are cut (i.e. whenever a bubble is cut);
the 8-loop graph (b) is not a tree bubble graph as it contains a loop of loops,
and it is not disconnected by the cut of a bubble in the loop.}
\end{center}
\end{figure}

In order to show the equivalence, let us write the
explicit interaction ${\cal L}_{int}$ that according to Eq.(\ref{Ldef}) and
neglecting constant terms becomes
\BE
{\cal L}_{int}(h)=\left(m_B^2+\frac{\lambda_B \varphi_0^2}{3!}\right)\varphi_0 h+
\frac{1}{2}\left(m_B^2-\Omega_0^2+\frac{\lambda_B \varphi_0^2}{2}\right)h^2
+\frac{1}{3!}\varphi_0\lambda_Bh^3
+\frac{\lambda_B}{4!}h^4.
\label{Lint} 
\EE 
The tree bubble graphs that contribute to the 1PI 2-point function
$\Gamma^\prime_2$ are shown in Fig.2 where a straight line represents the free propagator
$G_0=(p^2+\Omega_0^2)^{-1}$ as derived from the definition Eq.(\ref{Lgep}) of
${\cal L}_{GEP}$ at the vacuum $\varphi=\varphi_0$. 
The first order contribution $\Gamma_2^{(1)}$ arises from the first two graphs
of Fig.2 
\BE
\Gamma_2^{(1)}=\Omega_0^2-m_B^2-\frac{\lambda_B}{2} 
\varphi_0^2-\frac{\lambda_B}{2} I_0(\Omega_0)
\EE
and by insertion of the gap equation Eq.(\ref{gap}) we get $\Gamma_2^{(1)}=0$ 
which is a well known property of the GEP. Thus $\Gamma^\prime_2$ is given by
the sum of all the higher order tree bubble graphs shown in Fig.2.

\begin{figure}[htb]
\includegraphics[scale=0.7,bb=100 610 574 691,clip]{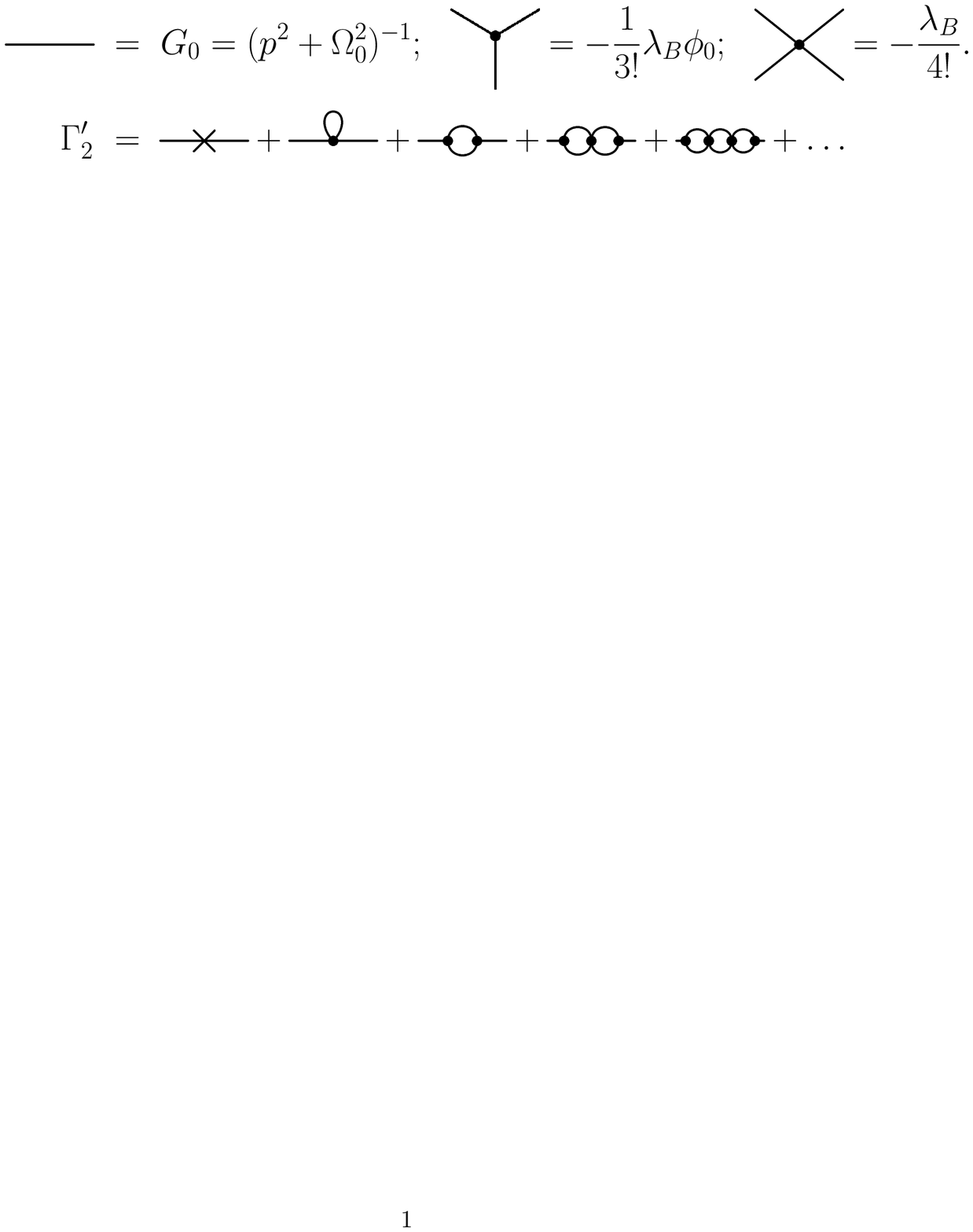}
\renewcommand{\figurename}{Fig.}
\caption{\label{Fig2} Tree bubble graphs contributing to the 1PI 2-point
function. 
Explicit expressions for the free propagator and for the vertices are reported
on the top.}
\end{figure}

Let us denote by $L_n$ the n-order 1-loop graph without vertex factors, 
without external lines and with external momenta set to zero, as shown in Fig.3.

\BE
L_{n+1}=\int_\Lambda \frac{d_E^4p}{(2\pi)^4}\frac{1}{(p^2+{\Omega_0}^2)^{n+1}}=
\frac{1}{n!}\left\vert \frac{d^n I_0}{d(\Omega^2)^n}\right\vert_{\Omega=\Omega_0}
\label{Ln}
\EE
and let us denote by $B$ the bubble chain geometric expansion reported as an hatched
bubble in Fig.4
\BE
B=\sum_{n=0}^{\infty} \left[-\frac{\lambda_B}{2}L_2\right]^n=
\frac{1}{1+\frac{\lambda_B}{2} L_2}.
\label{B}
\EE

\begin{figure}[htb]
\includegraphics[scale=0.6,bb=99 607 576 684,clip]{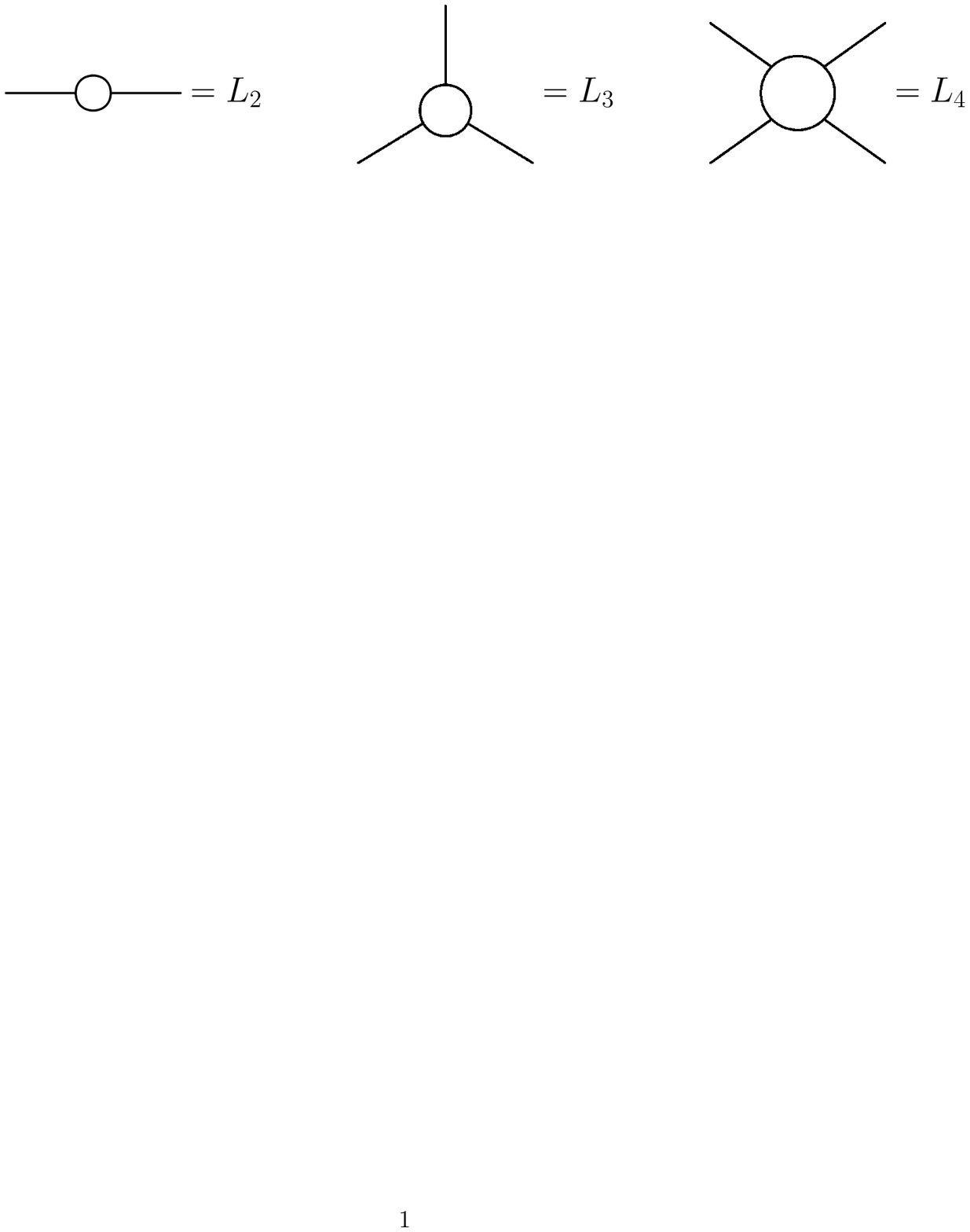}
\renewcommand{\figurename}{Fig.}
\caption{\label{Fig3} Graphs for the integrals $L_2$, $L_3$ and $L_4$.
They are n-order 1-loop graphs without vertices, 
without external lines and with external momenta set to zero.}
\end{figure}

\begin{figure}[htb]
\includegraphics[scale=0.7,bb=99 360 563 656,clip]{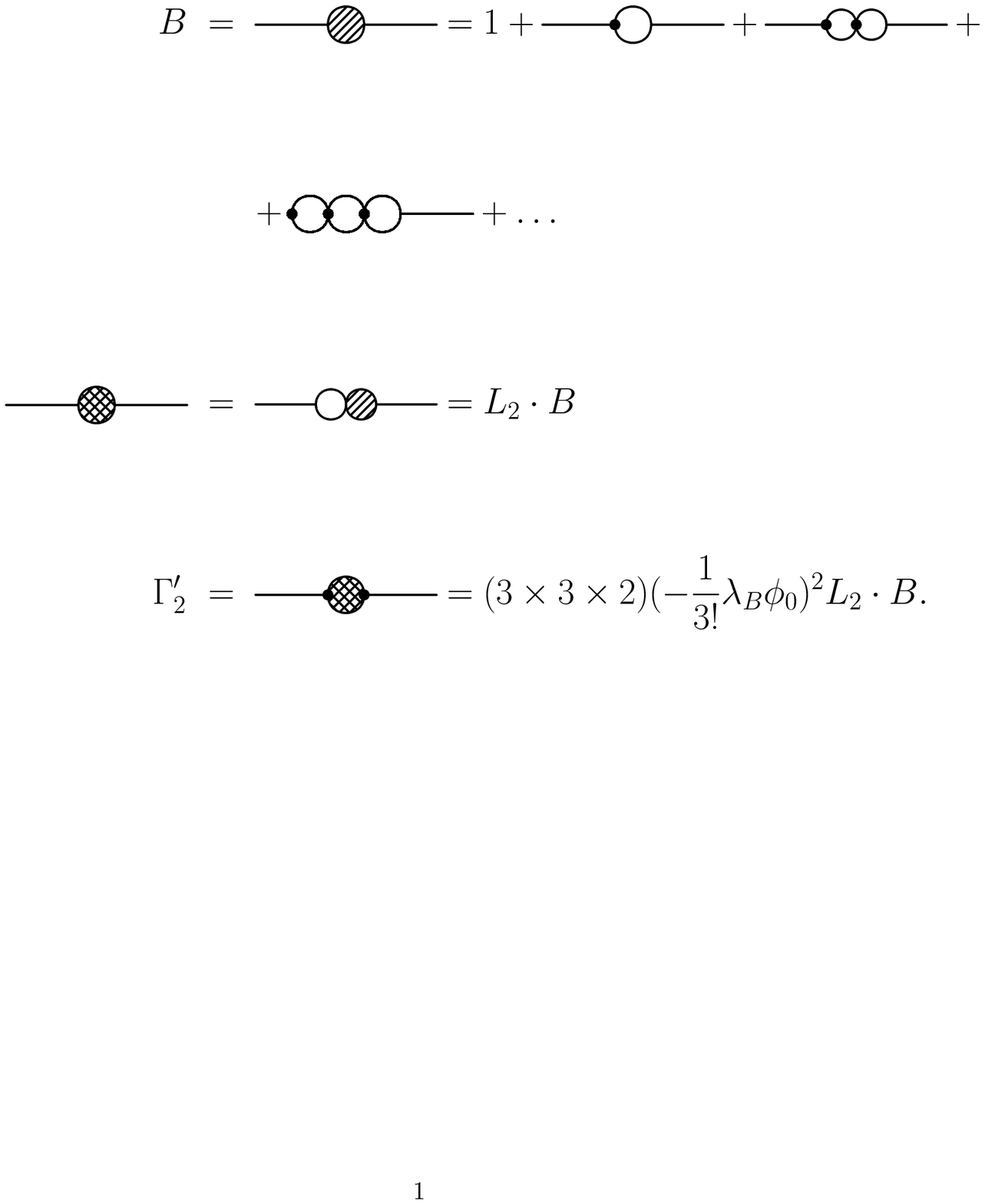}
\renewcommand{\figurename}{Fig.}
\caption{\label{Fig4} The bubble chain geometric expansion denoted by $B$ in the
main text (hatched bubble). The cross-hatched bubble represents the sum of all
the bubble chain graphs without external vertices.
The coupling $\Gamma_2^\prime(0)$ is obtained by adding the external vertices and
the correct symmetry factors to a cross-hatched bubble.  
}
\end{figure}

With the above notation the approximate 2-point coupling $\Gamma^\prime_2$ is shown in
Fig.4 as a cross-hatched bubble with vertex and symmetry factors added at the
external points, while the simple cross-hatched bubble represents the chain bubble
sum without external vertex factors. Accordingly we can write
\BE
\Gamma^\prime_2(0)=(3\times 3\times 2)\left(-\frac{1}{3!}\lambda_B \varphi_0\right)^2 
L_2 B
=\frac{3\Omega_0^2}{2}\lambda_B L_2 B
\label{G2}
\EE
where we have made use of Eq.(\ref{tree}). The renormalized mass $M^2_R=-\Gamma_2$
then follows
\BE
M_R^2=\Omega_0^2-\Gamma^\prime_2(0)=
\Omega_0^2\left[\frac{1-\lambda_B L_2}{1+\frac{\lambda_B}{2} L_2}\right].
\label{MR}
\EE

On the other hand, the derivative of Eq.(\ref{V2}) yields
\BE
\frac{d^2 V_{GEP}}{d\varphi^2}=(\Omega^2-\frac{1}{3}\lambda_B\varphi^2)
+2\varphi^2\left(\frac{d\Omega^2}{d\varphi^2}-\frac{\lambda_B}{3}\right)
\label{d2V}
\EE
and, at $\varphi=\varphi_0$, by insertion of Eq.(\ref{tree}) we get
\BE
\left[\frac{d^2 V_{GEP}}{d\varphi^2}\right]_{\varphi=\varphi_0}=
\frac{6\Omega_0^2}{\lambda_B}
\left[\left(\frac{d\Omega^2}{d\varphi^2}\right)_{\varphi=\varphi_0}-
\frac{\lambda_B}{3}\right].
\label{d2V0}
\EE
The derivative of $\Omega^2$ as a function of $\varphi^2$ can be easily
obtained by the gap equation Eq.(\ref{gap}) whose derivative reads
\BE
\frac{d\Omega^2}{d\varphi^2}=\frac{\lambda_B}{2}+\frac{\lambda_B}{2}
\left(\frac{d I_0}{d\Omega^2}\right) \frac{d\Omega^2}{d\varphi^2}.
\EE
Then the derivative of $\Omega^2$ is 
\BE
\frac{d\Omega^2}{d\varphi^2}=\frac{\lambda_B}{2}
\frac{1}{\left[1-\frac{\lambda_B}{2}\left(\frac{d I_0}{d\Omega^2}\right)\right]}
\label{domega}
\EE
which at the vacuum $\varphi=\varphi_0$ becomes
\BE
\left(\frac{d\Omega^2}{d\varphi^2}\right)_{\varphi=\varphi_0}=
\frac{\lambda_B}{2}
\frac{1}{\left[1+\frac{\lambda_B}{2} L_2\right]}.
\label{domega0}
\EE
Insertion of the derivative in Eq.(\ref{d2V0}) shows that, according to 
Eq.(\ref{gamma}) 
which defines the couplings as derivatives of the
GEP, Eq.(\ref{d2V0}) yields a renormalized
mass $M_R$ which is exactly the same as that obtained by the 
sum of tree bubble graphs in Eq.(\ref{MR}).

The equivalence can be extended to higher orders: the derivatives of the GEP
are easily evaluated by Eq.(\ref{d2V}) and give
\BE
g_R=\left[\frac{d^3 V_{GEP}}{d\varphi^3}\right]_{\varphi=\varphi_0}=
\varphi_0\lambda_B\frac{M_R^2}{\Omega_0^2}+4\varphi_0^3
\left[\frac{d^2\Omega^2}{d(\varphi^2)^2}\right]_{\varphi=\varphi_0}
\label{gR}
\EE
\BE
\lambda_R=\left[\frac{d^4 V_{GEP}}{d\varphi^4}\right]_{\varphi=\varphi_0}=
\lambda_B\frac{M_R^2}{\Omega_0^2}+
4!\varphi_0^2\left[\frac{d^2\Omega^2}{d(\varphi^2)^2}\right]_{\varphi=\varphi_0}+
8\varphi_0^4\left[\frac{d^3\Omega^2}{d(\varphi^2)^3}\right]_{\varphi=\varphi_0}
\label{lR}
\EE
where the derivatives of $\Omega^2$ follow from Eq.(\ref{domega}) and can
be written in compact form by insertion of Eq.(\ref{Ln}) and Eq.(\ref{B})
\BE
\left[\frac{d^2\Omega^2}{d(\varphi^2)^2}\right]_{\varphi=\varphi_0}=
\frac{\lambda_B^3}{4} B^3 L_3
\label{domega2}
\EE
\BE
\left[\frac{d^3\Omega^2}{d(\varphi^2)^3}\right]_{\varphi=\varphi_0}=
\frac{3}{8}\lambda_B^5 B^5 L_3^2-\frac{3}{8}\lambda_B^4 B^4 L_4.
\label{domega3}
\EE
With the above notation the renormalized couplings read
\BE
g_R=
\varphi_0\lambda_B\left[\frac{M_R^2}{\Omega_0^2}+(\varphi_0\lambda_B)^2 B^3L_3\right]
\label{gR2}
\EE
\BE
\lambda_R=
\lambda_B\left[
\frac{M_R^2}{\Omega_0^2}+6\varphi_0^2\lambda_B^2B^3L_3
-3\varphi_0^4\lambda_B^3 B^4 L_4
+3\varphi_0^4\lambda_B^4 B^5 L_3^2
\right].
\label{lR2}
\EE

\begin{figure}[htb]
\includegraphics[scale=0.7,bb=101 511 522 692,clip]{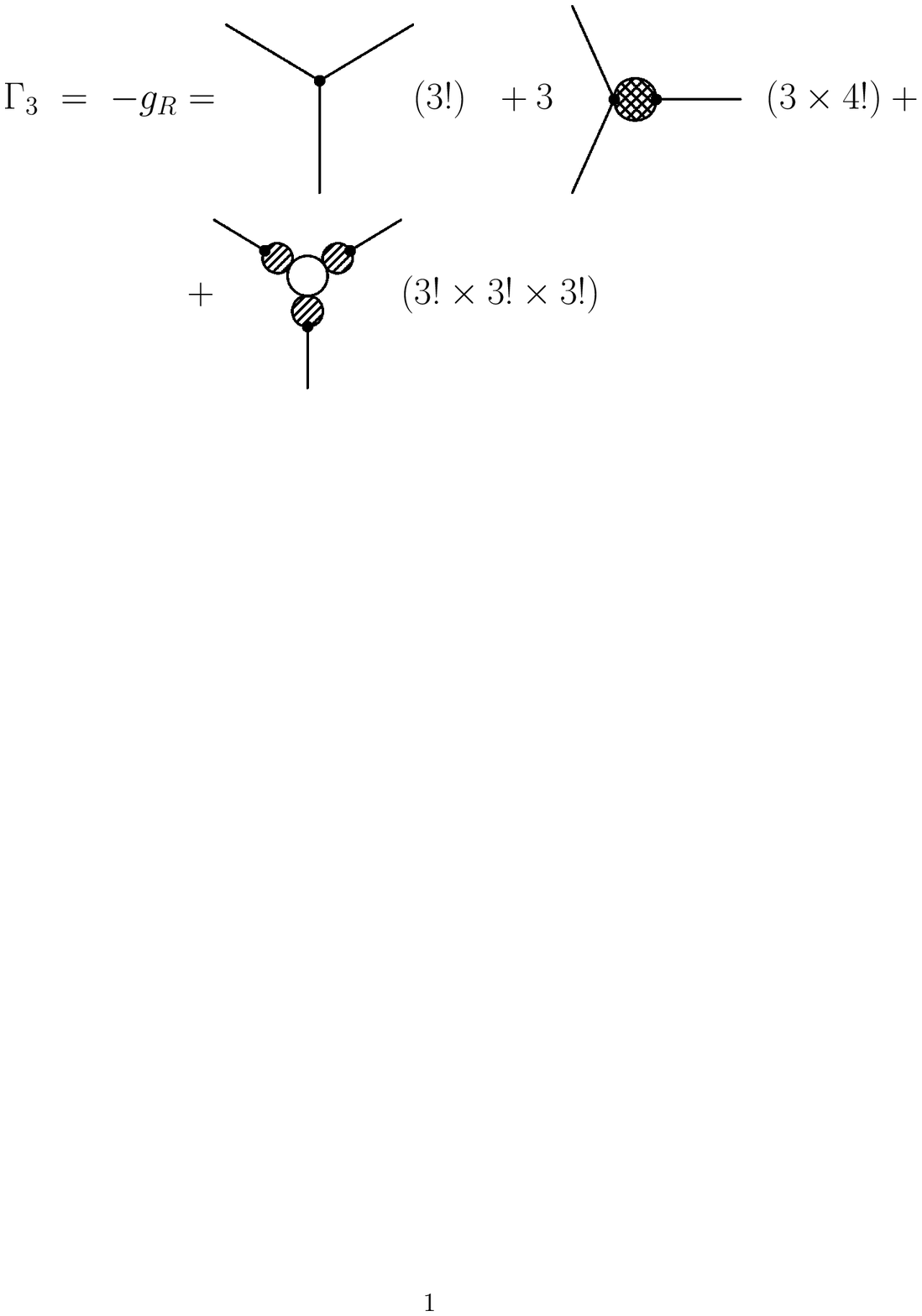}
\renewcommand{\figurename}{Fig.}
\caption{\label{Fig5} Tree bubble graphs contributing to the 1PI 3-point  vertex
function with their symmetry factors.}
\end{figure}

\begin{figure}[htb]
\includegraphics[scale=0.7,bb=100 333 569 693,clip]{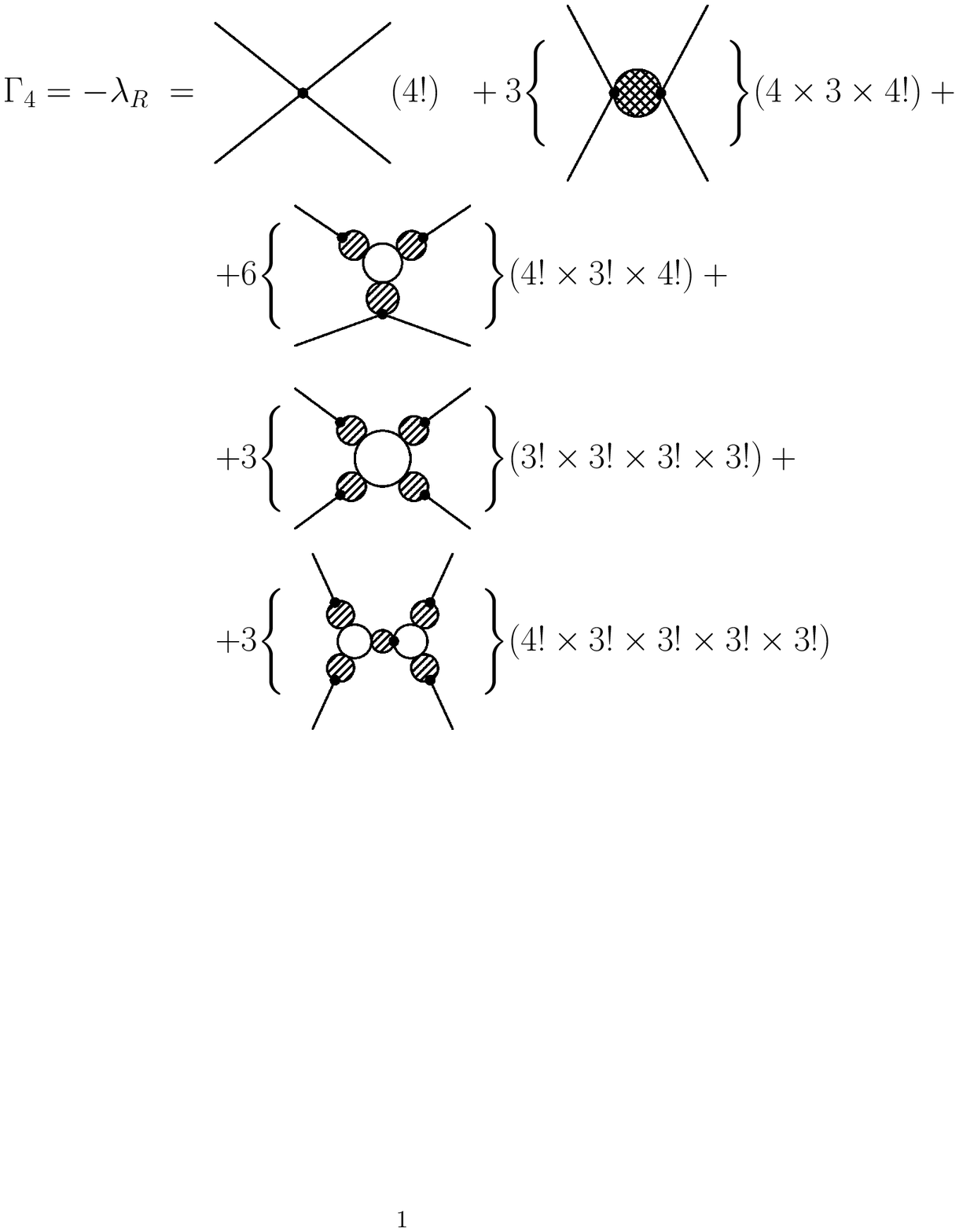}
\renewcommand{\figurename}{Fig.}
\caption{\label{Fig6} Tree bubble graphs contributing to the 1PI 4-point  vertex
function with their symmetry factors.}
\end{figure}

It is not difficult to show that these couplings can be recovered by the sum
of all the tree bubble graphs that can be drawn for the vertex functions. The
sum of the corresponding geometric expansions are reported in Fig.5 and Fig.6.
The tree bubble graphs that contribute to the 1PI 3-point vertex function are
reported in Fig.5 with their symmetry factors. The sum of the
first two terms is $\varphi_0\lambda_B-3\Gamma_2^\prime/\varphi_0
=\varphi_0\lambda_B M^2_R/\Omega_0^2$,
while the third graph yields $\lambda_B^3 \varphi_0^3 B^3L_3$, in agreement with
Eq.(\ref{gR2}). With the same notation the tree bubble graphs 
contributing to the 1PI 4-point vertex
function are reported in Fig.6: as for the 3-point function the sum of the first
two terms is $\lambda_B M^2_R/\Omega_0^2$ while the other graphs can be easily
shown to reproduce the right hand side of Eq.(\ref{lR2}) term by term.

The equivalence explains the content of the variational effective model
and makes it even more evident that the GEP provides a set of genuine 
non-perturbative approximate couplings that represent the sum to all 
orders of tree bubble graphs for the vertex functions.

\section{Renormalized couplings and Critical point}

Before we can compare the predictions of the model with the phenomenology, we
must address the problem of field renormalization.
As we already said, the variational mass $\Omega$ turns out to be quite
larger than the physical Higgs mass $M_h$, thus reducing any momentum
dependence of the n-point functions. As a consequence we expect a
negligible field renormalization and a physical mass $M^2_h$ which is
basically equal to $M^2_R=-\Gamma_2(0)$.
However, we would like to avoid any further approximation, and we prefer
to evaluate both the physical mass $M_h$ and the field renormalization factor
$Z_h$ from the 2-point function $\Gamma_2(p)$ according to the usual definitions:
\BE
\Gamma_2(p)\vert_{p^2=-M^2_h}=0
\label{renM}
\EE
\BE
Z_h^{-1}=-\frac{\partial \Gamma_2(p)}{\partial p^2}\vert_{p^2=-M^2_h}
\label{renZ}
\EE
Thus our finding that $Z_h\approx 1$ and $M_h\approx M_R$ can be regarded as
a check that the momentum dependence can be neglected, and a confirmation
that the 2-point function can be taken as
$-[\Gamma_2(p)]^{-1}=(p^2+M^2_h)$
as it was already assumed without proof
in previous work on the full gauge theory\cite{siringo_su2}.
The same conclusion had been reached by numerical simulations 
on the lattice\cite{kuti1,kuti2}.
Accordingly, the physical renormalized couplings must be defined as
\BE
\lambda_h=-Z_h^2\Gamma_4(0)=Z_h^2\lambda_R.
\EE
\BE
g_h=-Z_h^{3/2}\Gamma_3(0)=Z_h^{3/2} g_R.
\EE

The full 2-point $\Gamma_2(p)$ function may be recovered by the sum of the
bubble expansion in Fig.4 with the loop $L_2$ replaced by the function $L_2(p)$,
defined by the same graph in Fig.3 with the external momentum
restored
\BE
L_2(p)=
\int_\Lambda \frac{d_E^4 k}{(2\pi)^4}\frac{1}{\left[(p+k)^2+{\Omega_0}^2\right]}
\frac{1}{\left[k^2+{\Omega_0}^2\right]}
\label{L2p}
\EE
Then according to Eq.(\ref{G2}) the 2-point function reads
\BE
-\Gamma_2(p)=(p^2+\Omega^2_0)-\frac{3\Omega^2_0}{2}
\frac{\lambda_B L_2(p)}{1+\frac{\lambda_B}{2}L_2(p)}.
\label{G2p}
\EE
The same result can be found by the 
covariant Gaussian approximation\cite{kovner2}, 
or by direct functional differentiation of the Gaussian effective
action\cite{ibanez1},
and can be shown to be a genuine variational bound
of the exact 2-point function\cite{kovner1,ibanez1}. 

According to the definitions in Eqs.(\ref{renM}),(\ref{renZ}) the
physical Higgs mass $M_h$ is obtained as a solution of the
equation
\BE
M^2_h=\Omega^2_0\left[
\frac{1-\lambda_B L_2(p)}{1+\frac{\lambda_B}{2}L_2(p)}
\right]_{p^2=-M^2_h} 
\label{Mh}
\EE
while the field renormalization constant is given by
\BE
Z_h^{-1}=1-\frac{3}{2}\Omega^2_0\lambda_B\left[
\frac{\partial L_2(p)}{\partial p^2}
\left(1+\frac{\lambda_B}{2}L_2(p)\right)^{-2}
\right]_{p^2=-M^2_h}.
\label{Z}
\EE

The integral in Eq.(\ref{L2p}) must be evaluated inside
the four-dimensional hyper-sphere $k<\Lambda$, so that
the usual Feynman formula cannot be used. However, by
a tedious calculation the integral can be shown to yield
the following exact result
\BE
L_2(p)=\frac{1}{32\pi^2 p^2}
\left[\Lambda^2+p^2\log\frac{\Omega^2_0+\Lambda^2}{\Omega^2_0}-I(p)\right]
\label{L2pexact}
\EE
where the integral $I(p)$
\BE
I(p)=\int_{\Omega^2_0}^{\Omega^2_0+\Lambda^2}
\frac{dx}{x}\sqrt{(x-p^2)^2+4p^2\Omega^2_0}
\EE
is given by
\BE
I(p)=
p^2\left[t_1(t_2-t_1)
-\log\left\vert\frac{t_2}{t_1}\right\vert
+\left(\frac{t_1}{t_2}-1\right)
-\sqrt{1+4t^2_1}\>
\log\left\vert\frac{(t_2-t_-)(t_1-t_+)}{(t_2-t_+)(t_1-t_-)}\right\vert
\right]
\label{Ip}
\EE
with the variables $t_i,t_\pm$ defined according to
\BE
t_\pm=\frac{1}{2\Omega_0}\left[\pm\sqrt{p^2+4\Omega^2_0}-p\right]
\EE
\BE
t_1=\frac{\Omega_0}{p}
\EE
\BE
t_2=\frac{1}{2p\>\Omega_0}\left[\Lambda^2-p^2+\Omega^2_0+
\sqrt{4p^2\Omega^2_0+(\Omega^2_0+\Lambda^2-p^2)^2}\right].
\EE

In spite of the appearence, the loop function $L_2(p)$ in Eq.(\ref{L2pexact}) 
has a logarithmic divergence only: in fact for $M_h,\Omega_0\ll \Lambda$ we
can write, up to order $1/\Lambda^2$ 
\BE
L_2(p)\approx\frac{1}{16\pi^2}\left\{1+2\log\left(\frac{\Lambda}{\Omega_0}\right)
+\sqrt{1+\frac{4\Omega^2_0}{p^2}}\>\log\left[
1+\frac{p^2}{2\Omega^2_0}-\sqrt{\frac{p^2}{\Omega^2_0}\left(
1+\frac{p^2}{4\Omega^2_0}\right)}\>\right]\right\}
\label{L2papprox}
\EE
yielding at the point $p^2=-M^2_h$
\BE
L_2(p)\vert_{p^2=-M^2_h}\approx\frac{1}{16\pi^2}
\left\{1+2\log\left(\frac{\Lambda}{\Omega_0}\right)
-\frac{\sqrt{4\Omega^2_0-M^2_h}}{M_h}\>
\arctan\left(\frac{M_h\sqrt{4\Omega^2_0-M^2_h}}{2\Omega^2_0-M^2_h}\right)
\right\}
\label{L2M}
\EE
We notice that as far as $\Lambda$ is finite, large but not too large compared to
$\Omega_0$, the constant terms must be retained as they can be of the same
order as the logarithm $\log(\Lambda/\Omega_0)$.
From Eq.(\ref{L2M}) we see that $L_2(iM_h)$ is analytical in the limit
$M_h\to 0$ and its limit value $L_2(0)$ is
\BE
L_2(0)=\frac{1}{16\pi^2}\left[2\log\left(\frac{\Lambda}{\Omega_0}\right)-1\right]
\label{L20}
\EE
which is exactly the same large $\Lambda$ behaviour which would come out from
a direct calculation through the definitions of $L_n$ and $I_0$ by
Eqs.(\ref{Ln}),(\ref{I0}).

The limit $M_h\to 0$ is very important as it turns out to be the critical point
of the Higgs sector, since the vanishing of the mass may be associated to
a change of sign for the second derivative of the effective potential at its
stationary point which becomes a maximum. 
The exact equation for $M_h$, Eq.(\ref{Mh}), can be easily 
seen to admit the solution $M_h=0$ whenever the condition
\BE
\lambda_B =\frac{1}{L_2(0)} 
\label{auton}
\EE
is fulfilled. Thus we observe that, for a large but finite cut-off, 
the bare self-coupling $\lambda_B$
does not need to be small at the critical point: assuming that $\Omega_0/\Lambda$
is small enough we can make use of Eq.(\ref{L20}) and write the critical
condition Eq.(\ref{auton}) as
\BE
\Omega_0=\frac{\Lambda}{\sqrt{e}}\> e^{-\frac{8\pi^2}{\lambda_B}}.
\label{Omcrit}
\EE
We can check that for a coupling as large as $\lambda_B\approx 8\pi^2\approx 80$
we still find $\Omega_0/\Lambda\approx 0.22$ which is small enough to
be consistent with the use of the approximate Eq.(\ref{L20}) for $L_2(0)$.

From Eq.(\ref{Omcrit}) we see that there is no way to keep $\Omega_0$ finite
when the cut-off $\Lambda$ is sent to infinity unless the bare coupling
$\lambda_B$ is taken to be infinitesimal. This is just what 
happens in the autonomous
renormalization\cite{stevenson} 
where the bare self-coupling is taken to be $\lambda_B=1/L_2(0)$
while $L_2(0)$ diverges logarithmically. That is exactly the required condition
Eq.(\ref{auton}) for the vanishing of the Higgs mass. Thus it is not a great surprise
that in autonomous
renormalization the GEP predicts a vanishing Higgs mass\cite{ibanez1}.

From Eq.(\ref{Omcrit}) we also see that the GEP predicts a weak
first order transition because $\Omega_0$ does not vanish at the critical point.
According to Eq.(\ref{tree}) at the minimum of the effective potential
the vacuum expectation value of the scalar field is
$\varphi_0^2=3\Omega^2_0/\lambda_B$, and if $\Omega_0$ does not vanish at the
transition, the expectation value $\varphi_0$ jumps to zero at the critical
point. That seems to be an unavoidable shortcoming of the GEP, since
the transition is believed to be continuous. However by Eq.(\ref{Omcrit})
we see that the jump 
$\Delta \varphi$ of the expectation value at the transition is small
\BE
\Delta \varphi=\frac{\sqrt{3}\Lambda}{\sqrt{e\lambda_B}}
\> e^{-\frac{8\pi^2}{\lambda_B}}.
\label{jump}
\EE
It reaches its maximum at $\lambda_B=16\pi^2\approx 158$ where
$\Delta\varphi/\Lambda\approx 0.05$
and it goes to zero in both the limits $\lambda_B\to 0$ and $\lambda_B\to\infty$.
Thus we may neglect the jump in most cases, as far as we do not go too close
to the transition point. That is not a major problem at all, as we know that 
the physical range of interest cannot be too close to the critical point where
the Higgs mass eventually vanishes. We cannot use the GEP at criticality, but 
we expect that the order of the transition should not affect the behaviour
of the resulting effective model provided that we do not reach
the critical point. Actually we will see in the next section 
that the vacuum expectation value
$\varphi_0$ seems to be continuous up to the transition point for any reasonable
plot resolution.

Whenever $\Lambda\gg\Omega_0$ we can use the approximate Eq.(\ref{L2papprox})
for $L_2(p)$ in order to get the field renormalization constant $Z_h$.
Insertion of Eq.(\ref{L2papprox}) in Eq.(\ref{Z}) yields
\BE
Z_h^{-1}=1+\frac{\lambda_B}{96\pi^2} \frac{(2+x^2)^2}{(4-x^2)}
\left[1-\frac{32\pi^2\> f(x)}{x^2}\right]
\label{Zapprox}
\EE
where $x=M_h/\Omega_0$ and

\BE
f(x)=L_2(iM_h)-L_2(0)=\frac{1}{16\pi^2}\left[
2-\frac{\sqrt{4-x^2}}{x}\arctan\left(
\frac{x\sqrt{4-x^2}}{2-x^2}\right)\right].
\label{fx}
\EE 

We can simply check that $Z_h\approx 1$ in the small-coupling limit and
close to the critical point. That is quite obvious in the
small-coupling limit $\lambda_B\to 0$ where $M_h\approx\Omega_0$
and $x\to 1$. In this limit $Z_h^{-1}-1\approx 2\cdot 10^{-3}\cdot\lambda_B$,
which is vanishing small. In the opposite strong coupling limit we can
explore the critical range where $M_h\to 0$  ($x\to 0$) and, according
to Eq.(\ref{auton}), $\lambda_B L_2\approx 1$. In this limit
$96\pi^2f(x)\approx x^2$ and $Z_h^{-1}-1\approx \lambda_B/(144\pi^2)
\approx 0.7\cdot 10^{-3} \cdot\lambda_B$. This is just one per cent for a coupling
as large as $\lambda_B\approx 15$. In actual calculations we have never found 
values larger than few per cent in the broken symmetry phase, in agreement with
previous numerical findings\cite{kuti2,kuti1}.

Quite interesting, around criticality we find a hierarchy of energy scales with
an Higgs mass $M_h$ quite smaller than the mass parameter $\Omega_0$ which is
supposed to be much smaller than the energy cut-off $\Lambda$. The physical
meaning of the intermediate scale $\Omega_0$ is related to the vacuum
expectation value of the scalar field $v=247$ GeV according to Eq.(\ref{tree})
which reads $\Omega_0=v\sqrt{\lambda_B/3}$. For a strong coupling
$\lambda_B\approx 20$ we get $\Omega_0\approx 640$ GeV while $M_h$ can be
made as small as we like by a large cut-off.

\section{Renormalized couplings and Phenomenology}

\begin{figure}[htb]
\includegraphics[scale=2,bb=40 40 240 180,clip]{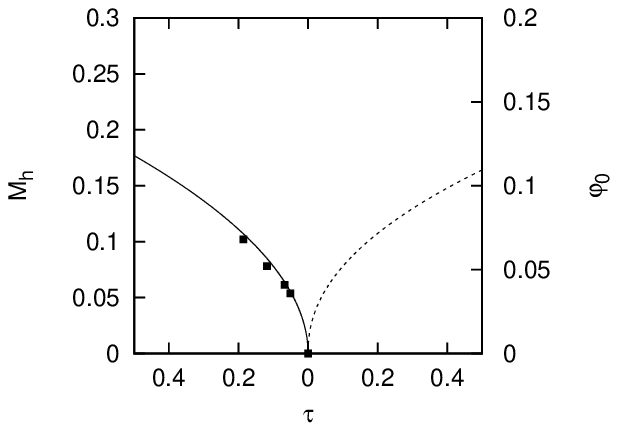}
\renewcommand{\figurename}{Fig.}
\caption{\label{Fig7} 
The physical Higgs mass $M_h$ (solid line) and
the vacuum expectation value $\varphi_0$ (dashed line) as functions of the adimensional
parameter $\tau=\vert 1-m_B^2/m_c^2\vert$, in the broken symmetry phase, 
for a moderately strong bare coupling $\lambda_B=10$. Data points are the
lattice simulations of Ref.\cite{oldMC1} for the symmetric phase, 
reported assuming $\Lambda a=4.38$.}
\end{figure}

In order to make contact with Monte Carlo calculations\cite{kuti2,kuti1} we
take all energies in units of the cut-off $\Lambda$. In these units the 
vacuum expectation value $\varphi_0=v/\Lambda$ can be seen as representing the
inverse of the cut-off in units of $v=247$ GeV. In fact at the critical point the
vanishing of $\varphi_0$ can be regarded as a restoration of symmetry at
a fixed cut-off, or as the effect of an infinite cut-off on a fixed vacuum
expectation value. Following Ref.\cite{kuti2} we take a constant
value for the bare self-coupling $\lambda_B$ and change the bare mass
$m_B^2$ up to the critical point which is reached when $M_h$ vanishes at
$m_B^2=m_c^2$. 
In the broken-symmetry phase we measure the distance from
the critical point by the adimensional parameter $\tau=\vert 1-m_B^2/m_c^2\vert$.
The physical Higgs mass $M_h$ and the vacuum expectation value 
$\varphi_0$ are reported in Fig.7 up to the transition point for a moderate 
$\lambda_B=10$. We observe
that the first-order jump of $\varphi_0$ is so small at the transition
that it cannot be even seen in the plot at any reasonable scale.

\begin{figure}[htb]
\includegraphics[scale=2,bb=40 40 240 180,clip]{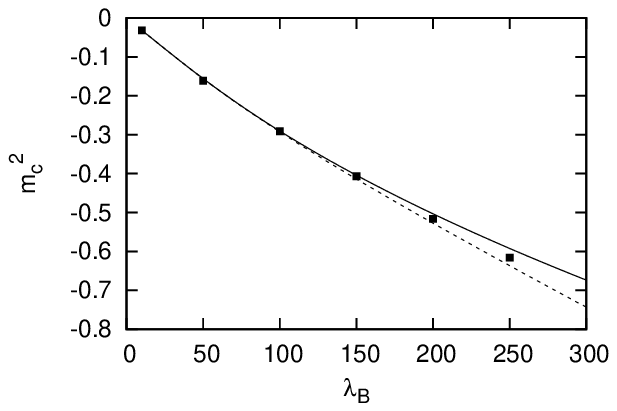}
\renewcommand{\figurename}{Fig.}
\caption{\label{Fig8} 
The critical parameter $m_B^2=m^2_c$ as a function of the
bare coupling $\lambda_B$. The solid line is the result of the present variational
method. The squares are the numerical lattice data
of Ref.\cite{oldMC1} scaled according to $\Lambda a=4.38$. 
The dahed line is the outcome of the 
approximate analytical function reported in Eq.(\ref{mcapp}).}
\end{figure}

A direct comparison with numerical simulations requires that the correct scale
factor should be fixed between energies. In lattice calculations energies are
usually taken in units of the inverse of the lattice spacing $a$ which also provides
a natural cut-off. Thus we expect that the scale factor $c=\Lambda a$ 
should be of order unity, but we have no direct and unique way to determine it.
For instance we could require that the unit cell of the inverse
lattice has the same extension of our four-dimensional hyper-sphere $p<\Lambda$,
and obtain $c=2^{5/4}\sqrt{\pi}\approx 4.2$. Here we prefer to make a more empirical
choice and take the scale factor $c$ as the ratio between the critical 
bare masses $m_c$, assuming that the critical masses should be the same 
if the energy scales are correctly handled. For a moderate $\lambda_B\approx 10-100$
there are not too many data to compare with: some numerical results have been reported
for the symmetric phase in Ref.\cite{oldMC1}, 
and a comparison of the critical bare masses
for $\lambda_B=10$ yields the empiric scale factor $c=4.38$. 

Once the scale factor
has been fixed we may compare our results with the lattice data. The critical point
$m^2_c$ is shown as a function of $\lambda_B$ in Fig.8 where the lattice data
of Ref.\cite{oldMC1} have also been included for comparison. Our result
fits the lattice data very well up to $\lambda_B\approx 150$. 
For larger couplings the GEP underestimates the strength of the critical $m^2_B$ as
a natural consequence of the predicted first order transition. On the other hand we
may extract from the GEP a yet simpler analytical approximate function that 
interpolates
the data very well for a moderate bare coupling. In fact
for a moderate
coupling $\lambda_B<100$ we already know that, according to Eq.(\ref{Omcrit}),
the ratio $\Omega_0/\Lambda$ is small enough at the
critical point. Therefore we can neglect powers higher than $\Omega_0^2/\Lambda^2$ in
the expansions of $L_2(0)$ and $I_0(\Omega_0)$, and inserting Eqs.(\ref{Omcrit}),
(\ref{tree}) in the gap equation Eq.(\ref{gap}) we gain the simple result
\BE
m^2_c=-\frac{\lambda_B}{32\pi^2}
\left[1-e^{-(1+\frac{16\pi^2}{\lambda_B})}\right].
\label{mcapp}
\EE
As shown in Fig.8, the approximate Eq.(\ref{mcapp}) gives the correct critical
point up to $\lambda_B\approx 200$, while for stronger couplings the strength
of the critical $m_B^2$ is overestimated.
We observe that Eq.(\ref{mcapp}) has an essential singularity at the point 
$\lambda_B=0$ and thus it cannot be recovered by any perturbative  
expansion in powers of $\lambda_B$. Despite being very simple, the analytical
result in Eq.(\ref{mcapp}) is a genuine non-perturbative prediction,
and provides an example of the capabilities of the variational method.

With the same scale factor, the available lattice data\cite{oldMC1} for the
mass 
have been inserted in Fig.7 for comparison, and again they are in good
agreement with the present variational calculation. Both calculations are 
consistent with a square root behaviour $M_h\sim\sqrt{\tau}$.
Furthermore, once the correct scale factor has been inserted, the numerical values
seem to be almost the same, with our data slightly larger than the lattice predictions.
That was not entirely expected as in the lattice simulation of Ref.\cite{oldMC1}
the renormalized mass
was measured in the symmetric phase while our results are for the broken symmetry
phase, and the critical behaviour is expected to agree in the two phases up
to constant factors: integration constants that are usually different in the
two phases. Some perturbative arguments have been reported in the past 
giving evidence for the equivalence of the integration constants 
in the two phases\cite{weisz},
while lattice data have shown\cite{kuti1} that, 
for a very strong coupling $\lambda_B=600$, the
constant factors are different, with the broken symmetry mass being almost twice 
the mass of the symmetric phase. Thus we argue that the slight difference
in Fig.7 can be taken as a measure of non-perturbative effects, still small for 
$\lambda_B=10$. Actually we have checked that when $\lambda_B$ increases the
difference also increases with the effect becoming quite large for $\lambda_B=50$ 
already. Unfortunately, as discussed at the end of the previous section, we cannot
allow $\lambda_B$ to become too large at the transition point since the transition
is predicted to be first order by the GEP. That precludes a direct comparison
with the lattice data of Ref.\cite{kuti1} in the broken symmetry phase.

\begin{figure}[htb]
\includegraphics[scale=2,bb=40 40 240 180,clip]{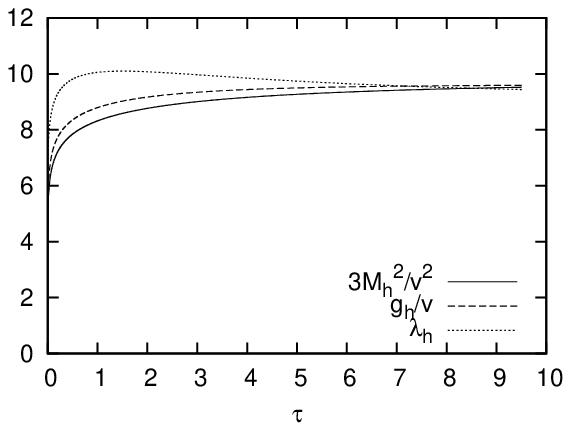}
\renewcommand{\figurename}{Fig.}
\caption{\label{Fig9} 
The physical adimensional renormalized couplings $3M^2_h/v^2$ (lower solid curve), 
$g_h/v$ (central dashed curve) and $\lambda_h$ (upper dotted curve)
as functions of the adimensional
parameter $\tau=\vert 1-m_B^2/m_c^2\vert$, in the broken symmetry phase, 
for a moderately strong bare coupling $\lambda_B=10$. The couplings merge in
the large $\tau$ limit (small cut-off).}
\end{figure}

\begin{figure}[htb]
\includegraphics[scale=2,bb=40 40 240 180,clip]{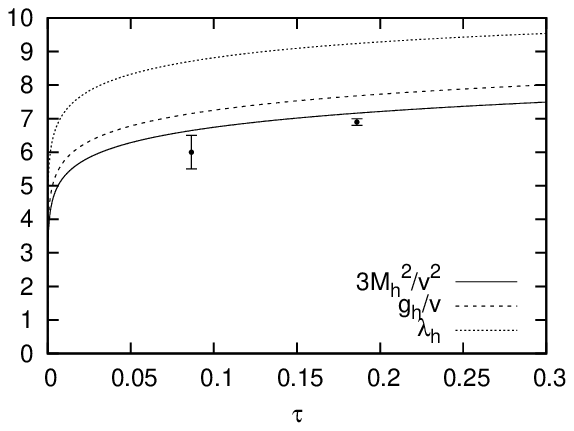}
\renewcommand{\figurename}{Fig.}
\caption{\label{Fig10} 
The physical adimensional renormalized couplings $3M^2_h/v^2$ (lower solid curve), 
$g_h/v$ (central dashed curve) and $\lambda_h$ (upper dotted curve)
as functions of the adimensional
parameter $\tau=\vert 1-m_B^2/m_c^2\vert$, in the critical range of the 
broken symmetry phase, for a moderately strong bare coupling $\lambda_B=10$.
Note the logarithmic decreasing of the couplings for $\tau\to 0$.
Data points are indirect lattice simulations for $3M^2_h/v^2$, and have been
obtained  by scaling from the symmetric phase data of Ref.\cite{oldMC1}.}
\end{figure}

In Fig.9 the physical adimensional renormalized couplings $3M^2_h/v^2$, $g_h/v$
and $\lambda_h$ are reported. For a large $\tau$ (i.e. a small cut-off) 
they all tend to the same bare value $\lambda_B$. 
At the critical point $\tau\to 0$ they all seem to vanish yielding a trivial model. 
However it should be observed that the renormalized coupling $\lambda_h$ is not
equal to $3M^2_h/v^2$ as it would be suggested by the standard tree-level
relation. Such a condition can only be satisfied
if $M_h\approx\Omega_0$ in Eq.(\ref{tree}), i.e. for a very small bare 
self-coupling or an unphysically small cut-off (large $\tau$). The same
adimensional renormalized couplings
are reported in Fig.10 for the critical range $\tau< 1$. For comparison
in the same Fig.10 some data points from Ref.\cite{oldMC1}
are reported as an indirect measure of $3 M_h^2/v^2$ on the lattice.
According to scaling, these points can be assumed to be equal to 
the value of $\lambda_R$ in the symmetric phase. In fact
for $\lambda_R$ we could not find any direct simulation data 
in the broken symmetry phase, nor any direct lattice measure
of the renormalized couplings $\lambda_h$ and $g_h/v$.

For small values of the parameter $\tau$ (i.e. for a large cut-off) all
the couplings decrease and show a logarithmic behaviour. 
We observe
that $\lambda_h$ is always larger than its perturbative value $3M^2_h/v^2$, 
yielding a strongly coupled effective model even when the Higgs mass falls 
well below the electrowek energy scale $v$. 
Moreover
the 3-point adimensional coupling $g_h/v$ and the 
4-point vertex $\lambda_h$ turn out to be different at variance with 
the case of a weak coupling where they are the same.
For a very light Higgs (close to the critical point), we may assume 
the autonomous condition Eq.(\ref{auton}),
take $B\approx 2/3$ in Eqs.(\ref{gR2}),(\ref{lR2})
and for a large cut-off write the ratio $g_h/(\lambda_h v)$ as
\BE
\frac{g_h}{v\lambda_h}\approx\frac{1}{4+\frac{\lambda_B}{8\pi^2}}
\label{glratio}
\EE
We observe that the ratio between the couplings is
$g_h/(v\lambda_h)\approx 0.25$ for any moderate bare coupling, and still
smaller if $\lambda_B$ gets very large, to be compared with the weak coupling
limit $g_h/(v\lambda_h)\approx 1$. 
The weakening of the 3-point coupling would give less chances of
finding bound states in the Higgs sector, since it is the 3-point vertex 
that can give a binding contribution in a strongly self-coupled
Higgs sector\cite{siringo_var,rupp}.

\section{Discussion}
The effective model that emerges variationally from the GEP
could provide a reliable way to describe the experimental
data, together with numerical lattice simulations. In fact we have shown that the
non-perturbative predictions of the GEP can be as effective as numerical
simulations, provided that the comparison is not pushed too close to the critical 
point where the GEP is known to fail. That precludes the study of extremely strong
couplings $\lambda_B\approx 100-1000$ as according to Eq.(\ref{jump}) the jump
of the vacuum expectation value $\varphi_0$ would not be negligible at 
the transition point. Unfortunately that has not allowed us to compare our
predictions with some numerical data\cite{kuti1} on the broken-symmetry phase
that have been reported for $\lambda_B=600$.

On the other hand most of the numerical simulations that have been reported for
moderate couplings $\lambda_B\approx 10-100$ happen to be too far from the
relevant physical range of parameters 
as the cut-off is usually too small\cite{oldMC1,oldMC2}. 
That represents
a common problem of numerical simulations since the finite size of the sample limits
the accuracy when the correlation length $\xi$ reaches the size $L$ of the sample.
Usually $L=na$ where $a$ is the lattice constant and $n$ is a small integer of order
ten at most, therefore taking $M_h=1/\xi$ and $\Lambda\approx 1/a$ we get a bound
$\Lambda<nM_h$. For a light Higgs $M_h\approx 100$ GeV and for $n=10$ we get a 
cut-off $\Lambda< 1$ TeV, too small for exploring the Higgs sector.
Actually, by a variational argument,
a threshold has been predicted at $\Lambda\approx 3.5$ TeV for the existence
of a strongly interacting light Higgs\cite{siringo_var}.

However some simulation data have been reported for the symmetric phase and a moderate
strong coupling\cite{oldMC1}, and have been found in perfect agreement 
with the predictions of the present method.
Larger lattices should be studied for a comparison in the physical
interesting range of parameters, while the real experiments seem to be 
an interesting and viable way to test the predictions of the present effective model.

It must be mentioned that before 
a full comparison can be made between the real phenomenolgy and
the effective model, the couplings to fermions and gauge fields must
be included in the simple scalar theory. Inclusion of gauge fields is
not a major issue as we have already derived the GEP for the full
$SU(2)\times U(1)$ non-Abelian gauge theory\cite{siringo_su2}, and
shown that the weak gauge couplings do not play any relevant role
in the Higgs sector. The couplings with fermions are expected to be
more important and, since they are proportional to masses, the Top quark
should be considered at least. The present variational method
can be extended to the Higgs-Top model that has been recently studied by
the GEP\cite{siringo_LR} and by numerical lattice simulation\cite{kuti2}.
While we do not expect any dramatic change of the general phenomenology with
respect to the present simple theory,
the resulting effective model should give a more reliable description of
the Higgs sector.

\end{document}